\documentclass[%
amsmath,amssymb,
aps,
]{revtex4-2}
\usepackage{graphicx}
\usepackage{dcolumn}
\usepackage{bm}
\usepackage{amsmath}
\usepackage{epstopdf}
\usepackage{hyperref}
\usepackage{color}

\begin{document}
 \title{Mutual coherent structures for 
 heat and angular momentum transport\\ 
 in turbulent Taylor-Couette flows}

 \author{ X.-Y. Leng}
 \author{J.-Q. Zhong}%
 \email{jinqiang@tongji.edu.cn}
 \affiliation{School of Physics Science and Engineering, \\Tongji 
  University, 200092 Shanghai, China}%

\date{\today}

\begin{abstract}
 
 In this study 
 we report numerical results of 
 turbulent transport of heat $Nu$ and 
 angular momentum $\nu_{t}/\nu$
 in Taylor-Couette (TC) flows 
 subjected to a radial temperature gradient.
 Direct numerical simulations are performed in a TC cell
with a  radius ratio $\eta=0.5$ and 
 an aspect ratio $\Gamma=8$
for two Rayleigh numbers ($Ra=10^5$, $10^6$) and
 two Prandtl numbers ($Pr=0.7$, $4.38$),
 while the Reynolds number $Re$ varies 
 in the range of $0 \le Re \le 15000$.
 With increasing $Re$, 
 the flows undergo two distinct transitions: 
 the first transition 
 being from the convection-dominated regime to the transitional regime,
 with the large-scale meridional circulation 
 evolving into spiral vortices; 
 the second transition occurring in the rotation-dominated regime
 when  
 Taylor vortices turn from a weakly non-linear state into a turbulent state.  
 In particular,
  when the flows are governed by turbulent Taylor vortices,
 we find that
 both transport processes exhibit power-law scaling: 
 $Nu\sim Re^{0.619\pm0.015}$ for $Pr=4.38$,
 $Nu\sim Re^{0.590\pm0.025}$ for $Pr=0.7$ and 
 $\nu_{t}/\nu\sim Re^{0.588\pm0.036}$ for both $Pr$.
These scaling exponents
 suggest 
 an analogous mechanism for 
 the radial transport of heat and angular momentum,
 which is further evidenced by the fact that 
 the ratio of turbulent viscosity to diffusivity is independent of $Re$.
To illustrate the underlying mechanism of turbulent transport,
we extract the coherent structures 
by analyzing  the spatial distributions of 
heat and momentum flux densities. 
Our results reveal mutual turbulent structures 
through which both heat and angular momentum are transported efficiently.

\end{abstract}

\maketitle

\section{\label{sec:level1}Introduction }

Turbulent transport processes of heat, mass and momentum are the 
central aspects in studying turbulence, 
owing to their close relations to various 
natural flows
\cite{hussain_1986,bergman2011fundamentals,
 holmes_lumley_berkooz_rowley_2012,
grossmann2016high,ahlers2009heat}. 
To understand the mechanism of turbulent transport
 is a challenging task in 
fluid physics and  crucial for the related applications.
Taylor-Couette (TC) flow, 
a fluid layer driven by two concentrically rotating 
cylinders, 
is of fundamental interest in 
many perspectives 
\cite{fardin2014hydrogen,grossmann2016high}, 
for example, in probing the angular momentum transport in accretion 
disks \cite{ji2006hydrodynamic,Ji2013AngularMT,Avila_PRL}.
It is also relevant to various 
 applications in industry such as drag reduction
\cite{Berg_Drag_PRL_2005,Gils_JFM_drag} and
solidification \cite{VIVES1988,McFadden_PoF}.

In TC flows, 
the toroidal motion of Taylor vortex (TV) 
may enhance the mixing and transport efficiency.
Hence
 TC reactors are extensively applied to chemical, food and biology processes
\cite{kataoka1995emulsion,Haut_bioreactor,Masuda_starch}.   
In these applications,
 a heated or cooled cylinder is inevitable.
 It is thus desirable to investigate the flow structures and transport 
properties in the TC systems subjected to a radial temperature 
gradient.
For modest Reynolds number ($Re$), 
studies of flow regimes, instabilities and 
pattern formations in such TC systems
 have attracted a lot of attention,
including experiments
\cite{Snyder_64,sorour1979effect,BALL89,
 lepiller2008weak,Mutabazi2019,Wen2020controlling},
 stability analyses 
\cite{ali_weidman_1990,kang2015,kirillov_mutabazi_2017,Yoshikawa13}
and numerical simulations
\cite{Kedia98, Guillerm15,Teng2015,Kang2017}.
However, in turbulent TC flows,
much less effort has been made to investigate
 the complex problem of  
the turbulent transport processes,
which are supposed to be more relevant to most applications 
in geophysical and industrial flows
\cite{Ji2013AngularMT,Avila_PRL,Berg_Drag_PRL_2005}. 
To better understand the relationships between the 
scalar and momentum transport in high-$Re$ regime, 
it is crucial to predict the
 interior 
structures and states of the flows. 
Furthermore,
 determination of the scaling laws of heat and momentum transport
is vital to extrapolate the existing results from laboratories to
large-scale geo- and astrophysical flows.
It remains a challenging question to date 
whether the scalar and momentum transport by TC flows 
share similar scaling behaviors in the turbulent regime
\cite{lathrop1992transition,lathrop1992turbulent,Leng_2021}.

Coherent structures play an important role in turbulent transport 
processes \cite{holmes_lumley_berkooz_rowley_2012}.
In turbulent TC flows, 
the momentum transport is implemented by 
the coherent structures in forms of turbulent TVs
and turbulent plumes between  adjacent TVs
\cite{ostilla2014exploring,Veen2016Taylor,Froitzheim2019Angular}.  
The meridional advection of TVs sweeps the radial and axial boundaries 
simultaneously,
potentially  providing a similar transport mechanism in both 
directions.
Indeed, in recent simulations \cite{Leng_2021},
we find that 
when an axial temperature gradient is applied in turbulent TC flows, 
the axial heat-transport scaling is analogous to that of 
the radial transport of angular momentum \cite{Leng_2021}.
This result confirms 
the existence of analogy between the axial dispersion of a passive scalar 
and the radial transport of momentum
\cite{lathrop1992transition,lathrop1992turbulent}. 
In the scenario that 
a radial temperature difference is applied in TC systems, 
both heat and angular momentum can be transported radially. 
In this system,
 how the large-scale structures, 
 such as TVs, 
 affect the turbulent transport processes 
 is  a natural question of great interest.

In this study, 
we utilize  
the paradigmatic model of  TC systems consisting of 
a heating (cooling) inner (outer) cylinder with 
two adiabatic endwalls. 
We consider the radial
transport processes of angular momentum and heat 
 in a high-Reynolds-number regime.
The results suggest that,
in the regime of turbulent TVs,
the radial transport of heat and angular momentum 
possess similar scaling relationships.
Furthermore, by extracting  fluid domains of 
high flux densities,
we demonstrate  that 
the heat and momentum transport 
are manipulated mainly by similar turbulent structures.

\section{\label{sec:level1}Numerical simulations }

\subsection{\label{sec:level2}{Physical model}}

We investigate the  three-dimensional flow of an incompressible viscous 
fluid contained between two concentric cylinders of 
radii $r_1$, $r_2$ 
and height $h$. 
The inner wall is rotating about $\mathit{z}$ axis $(e_{z})$ with angular 
velocity $\omega_1$, while the outer one is set to be fixed. 
A radial temperature difference $\Delta$ is imposed on the cylinders 
with the hot inner ($t_1$) and cold outer ($t_2$) walls.
The fluid properties including 
kinematic viscosity $\nu$, 
thermal expansion coefficient $\beta$ and 
 thermal diffusivity $\kappa$ 
 are assumed to be constant.
The governing parameters are
 the Rayleigh number  
 $\mathit{Ra}=\beta{g}\Delta{d}^3/{(\nu\kappa)}$, 
the Prandtl number 
$\mathit{Pr}=\nu/\kappa$ 
and the Reynolds number $Re=\omega_1r_1d/\nu$ respectively,
where $d=r_2-r_1$ is the gap width
and $g$ is the gravitational acceleration. 
The Richardson number 
$\mathit{Ri}=Ra/Pr/Re^2$, 
defined as the ratio of the free fall 
velocity to the inner-wall velocity, 
is adopted here 
to measure the relative strength between thermal 
convection and TC flow.
Two important geometrical parameters entering into the problem
are the aspect ratio $\Gamma=h/d$ and the radius ratio $\eta=r_1/r_2$.
The gap width $d$, imposed temperature difference $\Delta$  
and inner-wall velocity $u_1=\omega_1r_1$ 
are introduced as the length, temperature and velocity scales. 
Therefore, within the Boussinesq approximation, 
the dimensionless Navier-Stokes equations are
\begin{equation}
 \frac{\partial\textbf{U}}{\partial{\tau}}+
 ({\textbf{U}}\cdot\nabla){\textbf{U}}=
 -\nabla{p}+{\frac{1}{Re}}{\nabla}^2\textbf{U}
 +RiT{\mathit{e_z}},
 \qquad\nabla\cdot{\textbf{U}}=0,
\end{equation} 
\begin{equation}
 \frac{\partial{T}}{\partial{\tau}}+({\textbf{U}}\cdot\nabla){T}=
 \frac{1}{{RePr}}{\nabla}^2{T},
\end{equation}
where $\mathit{\tau}$, $\mathit{p}$ and $\mathit{T}$ are, correspondingly, 
time, 
pressure and temperature. 
And $\textbf{U}$ ($\mathit{U_r}$, 
$\mathit{U_\theta}$, $\mathit{U_z}$) are the components of velocity in radial, 
azimuthal and axial directions for cylindrical coordinates ($\mathit{R}$, 
$\mathit{\Theta}$, $\mathit{Z}$) 
respectively. 
The lower case letters $t$,
$\textbf{u}$
($\mathit{u_r}$, $\mathit{u_\theta}$, $\mathit{u_z}$) and
($\mathit{r}$, $\mathit{\theta}$, $\mathit{z}$) 
denote the dimensional temperature, velocity and coordinates.
The dimensionless fluid angular velocity is  $\omega$.
It is demonstrated in
Appendix A that 
the effect of centrifugal buoyancy \cite{Yoshikawa13, Guillerm15}
 does not change the main results 
 and is thus neglected. 
The inner and outer cylinders are maintained 
at fixed temperatures $T_1=1$ and $T_2=0$ respectively, 
while endwalls are set to be thermally insulating. No-slip 
boundaries are applied for velocities at all walls.
We use a wide gap with the radius ratio $\eta=r_1/r_2=0.5$, 
and the aspect ratio is $\Gamma=h/d=8$. 
This small-$\eta$ system has been discussed 
widely by numerical simulations \cite{DONG2007Direct} 
and experiments \cite{Merbold2013,Veen2016Taylor}.

Heat and 
angular momentum 
are two important transport quantities in the present system.
In general,
the global heat and angular momentum transport
are expressed by 
 $Nu$ \cite{grossmann2000scaling}
and  $Nu_{\omega}$
\cite{eckhardt2007torque} respectively.
The Nusselt number is defined as 
$\mathit{Nu}=a^{-1}
({{\mathit{RePr}} 
 \left \langle U_rT\right\rangle_{V} 
 - \partial _{r}\left \langle 
 {T}\right\rangle_{V}})$, 
where $\left \langle \right\rangle_{V}$ 
denotes the volume- 
and time-averaging 
and the geometry factor $a=2d^2/(ln(r_2/r_1)(r_2^2-r_1^2))$
 is induced by the annular gap (see Appendix B).
However, owing to the braking effect of the fixed endwalls, 
$Nu_{\omega}$ decreases along the radial direction in our system 
\cite{Leng_2021}. 
As introduced in Appendix B, 
we use the dimensionless effective viscosity 
${\nu_{t}}/{\nu}={(2Re_\tau)^2}/{Re}$  to
 represent the angular momentum transport instead of  $Nu_{\omega}$.
 Here, $Re_\tau=0.5u_\tau d/\nu$ is the friction Reynolds number 
 with the friction velocity
 $u_\tau^2=-\nu r  (\partial\left\langle 
 u_{\theta}/r \right\rangle _{\theta z}/ \partial r)$
 at the inner wall,
 where $\left \langle \right\rangle_{\theta z}$ 
 denotes the azimuthal-, axial- and time-averaging.

\subsection{\label{sec:level2} Numerical method}

\noindent

The equation system is solved using the finite difference scheme 
developed in Ref. 
\cite{krasnov2011comparative} and modified 
for the cylindrical coordinate \cite{ leng2018numerical, 
 Akhmedagaev2020Turbulent,Leng_2021}.
The numerical scheme is a
second-order approximation based on the spatial discretization, which is nearly 
fully
conservative with regard to mass, momentum and kinetic energy. The second-order
explicit Adams–Bashforth/backward-differentiation scheme is employed for the 
time
discretization. The viscous terms are treated explicitly, and implicit 
treatment is applied
for the diffusion term. At every time step, two Poisson equations, the 
projection method
equation for pressure and the equation for temperature are solved using 
fast-Fourier
transforms in the azimuthal direction and the cyclic reduction direct solver 
\cite{adams1999efficient}.
Towards the walls, the clustered grid is implemented using the hyperbolic 
tangent coordinate 
transformation.

The grid sensitivity studies as well as the main results 
are listed in {Tables} \ref{tab1:Res},
 \ref{tab2:results_4.38} and \ref{tab3:results_0.7}
 in Appendix A.
For each set of $Ra$, 
results from previous low-$Re$ convection 
are used as the initial condition for 
the following high-$Re$ case.
Data from initial transient state are excluded, 
and data taken over statistically steady state 
are averaged to determine the 
heat and angular momentum transport.  
The time convergence is checked by comparing 
 the time averages over 
the whole and the last halves of the 
simulation, 
and the resulting discrepancy is less than $3\%$.
For the temporal resolution, 
the chosen time step $\Delta t$ satisfies the Courant-Friedrichs-Lewy (CFL) 
condition,
and the CFL number remains less than 0.5.
The total run time for each case 
(including the initial and the averaging stages) 
is  greater than 200 large eddy turnover time units, 
and the averaging time is not less than 
100 large eddy turnover time units.

\section{\label{sec:level1} Results and discussions}

\subsection{\label{sec:level2} Global transport of heat and angular momentum }

\begin{figure} [ht]
 \centering
 \includegraphics[width=6in]{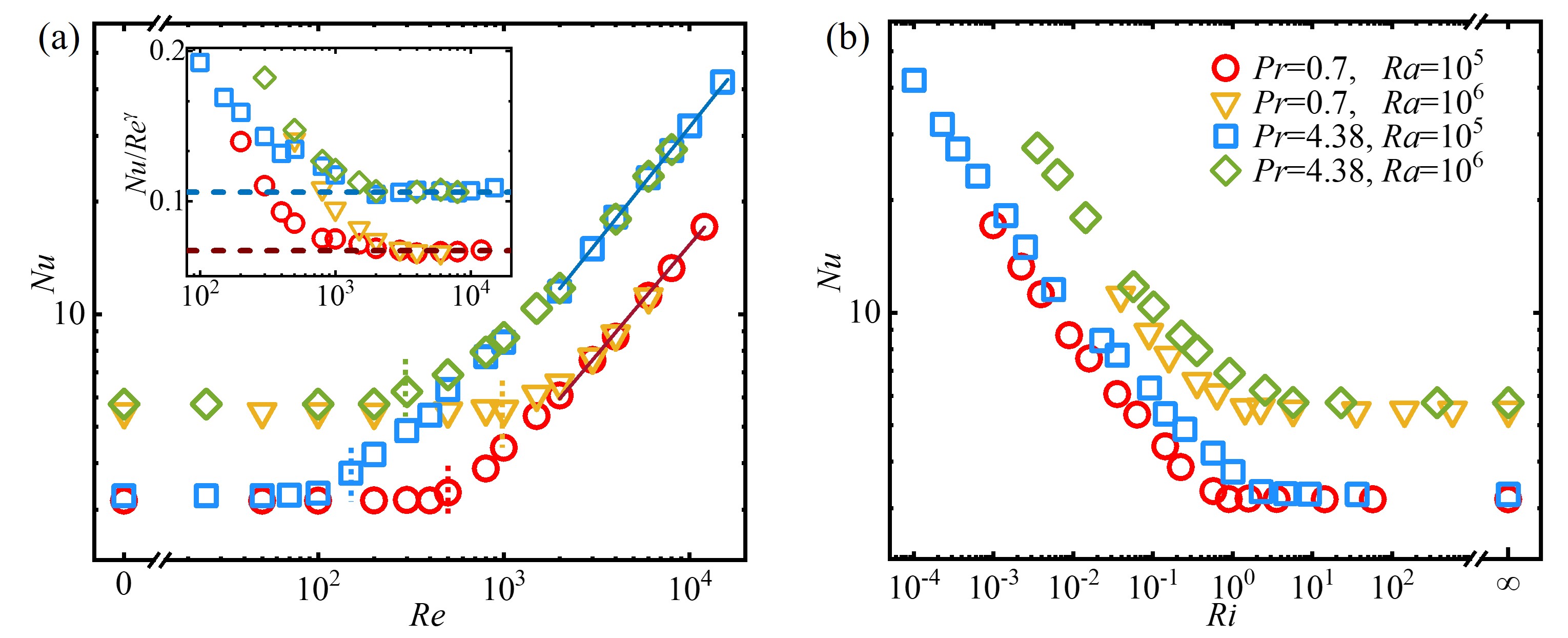}
  \caption{ \label{fig1}
  (a) Global heat transport $Nu$ as a function of $Re$ for $Pr=0.7$ and 
  $4.38$.
  Symbols are defined in panel (b).
  Solid lines denote the power-law fitting 
  $Nu= \alpha Re^{\gamma}$
   with $\gamma = 0.619$ for $Pr=4.38$ (blue) and 
   $\gamma = 0.590$ for $Pr=0.7$ (red).  
   Vertical dashed lines denote the first transition $Re^*_1$.
  Inset: an expanded view of the
  compensated plot of $Nu$ as a function of $Re$.   
  Vertical dotted lines denote the second transition $Re^*_2$.
  Dashed auxiliary lines indicate the values for the prefactor  
  $\alpha=Nu/Re^{\gamma}$=0.106 with $\gamma = 0.619$ (blue)  
  and 
  $\alpha$=0.068 with $\gamma = 0.590$ (red).  
   (b) $Nu$ as a function of $Ri$.
 }
\end{figure}

\subsubsection{Radial heat transport}

We first examine the global transport of heat. 
Data of the Nusselt number $Nu$ are 
shown as a function of $Re$ in 
{Fig.} \ref{fig1}(a) for 
 $Ra=(10^5$, $10^6)$ and
 $Pr=(0.7$, $4.38)$.
Without or with weak rotations 
($0\le Re \le 100$),
the flow and heat transfer are dominated 
by vertical convection \cite{Shishkina16PRE},
 and  $Nu$ is nearly independent of $Re$. 
We see that
in this buoyancy-dominant flow regime 
$Nu$ is larger for a greater $Ra$  for a given $Re$, 
but nearly independent of $Pr$.
With increasing $Re$,  
$Nu$ starts to grow when $Re$ 
exceeds a critical value $Re_1^*$
given  in {Table} \ref{tab3:Re_Ri}.
Interestingly, 
 $Nu$ for each  $Pr$ converges 
and becomes independent of $Ra$,
indicating the fading away of buoyancy-driven convection.
Further increasing $Re$,
a second transition takes place at $Re^*_2$,
after which $Nu$ exhibits a power-law scaling 
$Nu \sim Re^{\gamma}$ with 
$\gamma=0.590\pm0.025$
 for $Pr=0.7$ and
 $\gamma=0.619 \pm 0.015$
for $Pr=4.38$
(see the compensated plot in the inset of 
{Fig.} \ref{fig1}(a)).
Both scalings for $Pr=0.7$ and $4.38$ 
suggest the existence of 
a new flow regime for heat transfer.
All the transitional values $Re^*_1$ and $Re^*_2$
are listed in 
{Table} \ref{tab3:Re_Ri}. 
We find that at lower $Pr$ or higher $Ra$, 
the more vigorously convective flows
postpone the transitions.

We see in  {Fig.} \ref{fig1}(a) 
the intriguing trend that
 in a TC system the radial heat transport 
 $Nu$ is insensitive to the variation of $Pr$ 
 in a low-$Re$ flow regime ($Re\textless Re_1^*$), 
 but becomes strongly dependent on $Pr$
 (independent of $Ra$) with sufficiently high $Re$. 
 In the intermediate regime of $Re$, 
 our data curves of $Nu(Re)$ show complicated 
 $Ra$- and $Pr$-dependence. 
To better clarify the 
variations of the heat transport 
with changing control parameters, 
we show in {Fig.} \ref{fig1}(b) 
the Nusselt number as a function of  Richardson number $Ri$
which measures the relative strength of buoyancy and rotation.
In a high-$Ri$ regime where 
the buoyancy-driven convection is dominant 
(corresponding to the low-$Re$ regime shown in {Fig.} \ref{fig1}(a)), 
we see that $Nu$ remains a constant. 
The first transition for $Nu$-enhancement takes place at
$Ri_1^*$ that depends on both $Ra$ and $Pr$.
When $Ri \le Ri^*_1(Ra,Pr)$,
the strong rotations start to affect the flows
and $Nu$ increases monotonically as $Ri$ decreases.
In comparison with the complicated $Ra$- and $Pr$-dependence 
shown in {Fig.} \ref{fig1}(a),
a clear trend is displayed:
the curves of $Nu(Ri)$  collapse approximately 
for the same $Ra$ with various $Pr$,
but become well distinguishable 
when different $Ra$ is used. 
For a given $Ri$, $Nu$ increases with increasing $Ra$.

\begin{figure}[ht]
 \centering
 \includegraphics[width=6in]{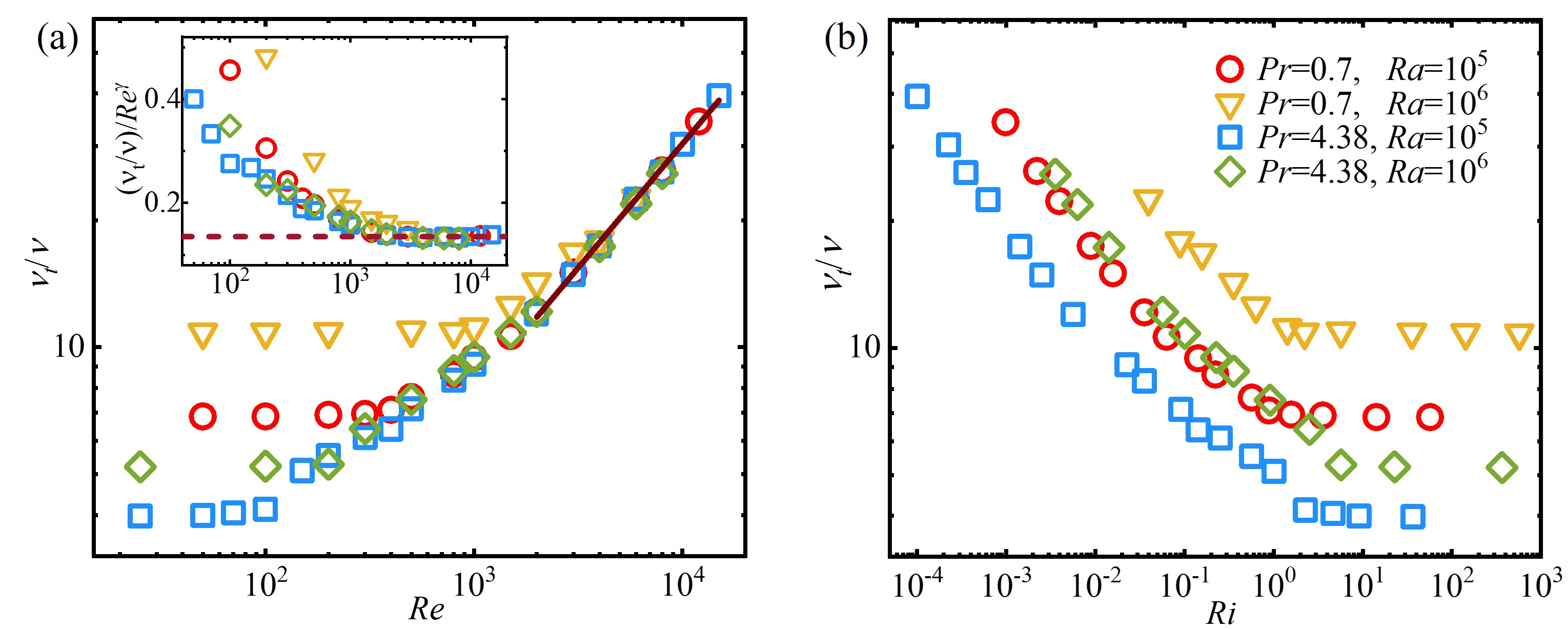}
 \caption{ \label{fig2} 
  (a) Dimensionless effective viscosity $\nu_{t}/\nu$
  as a function of $Re$ for $Pr=0.7$ and $4.38$.
  Symbols are defined in panel (b).
  Solid line denotes the power-law fitting 
  $\nu_{t}/\nu= Re^{\gamma}$ with ${\gamma}=0.588$.
  Inset:  an expanded view of the compensated plot of $\nu_{t}/\nu$ as a function of $Re$. 
  Dashed auxiliary line indicates the value for the prefactor  
  $\alpha=(\nu_{t}/\nu)/Re^{\gamma}$=0.135 with $\gamma = 0.588$.
  (b) $\nu_{t}/\nu$ as a function of $Ri$.
 } 
\end{figure}

\subsubsection{Angular momentum transport}

{Figure} \ref{fig2}(a)
presents the dimensionless effective viscosity $\nu_{t}/\nu$
 as functions of $Re$ for various $Pr$ and $Ra$, 
 which reveals the global angular momentum transport. 
In a low-$Re$ regime 
where vigorous convection governs the flows, 
$\nu_{t}/\nu$ is independent of $Re$, 
but increases with a larger $Ra$ or with a smaller $Pr$. 
Analogous to the heat transport data, 
we see that  $\nu_{t}/\nu$ starts to increase 
when $Re$ exceeds the transitional value $Re_1^*(Ra,Pr)$
obtained in {Fig.} \ref{fig1}(a). 
For various $Ra$ and $Pr$,
data of $\nu_{t}/\nu(Re)$
converge to the same curve for $Re\textgreater Re_1^*$,
and follow
a unifying power-law scaling 
$\nu_{t}/\nu\sim Re^{0.588\pm0.038}$ 
after the second transition
for $Re\textgreater Re^*_2$
(see the compensated plot in inset). 
In {Fig.} \ref{fig2}(b)
where the data are plotted as a function of $Ri$, 
we see that with decreasing $Ri$,
  $\nu_{t}/\nu$ starts to increase 
at $Ri^*_1$.
Unlike the converging of 
$Nu(Ri)$ for a given $Ra$ and with varying $Pr$ 
as shown in {Fig.} \ref{fig1}(b),
curves of $\nu_t/\nu(Ri)$ are found to be dependent 
on both $Ra$ and $Pr$.

\begin{figure} [htbp]
 \centering
 \includegraphics[width=3in]{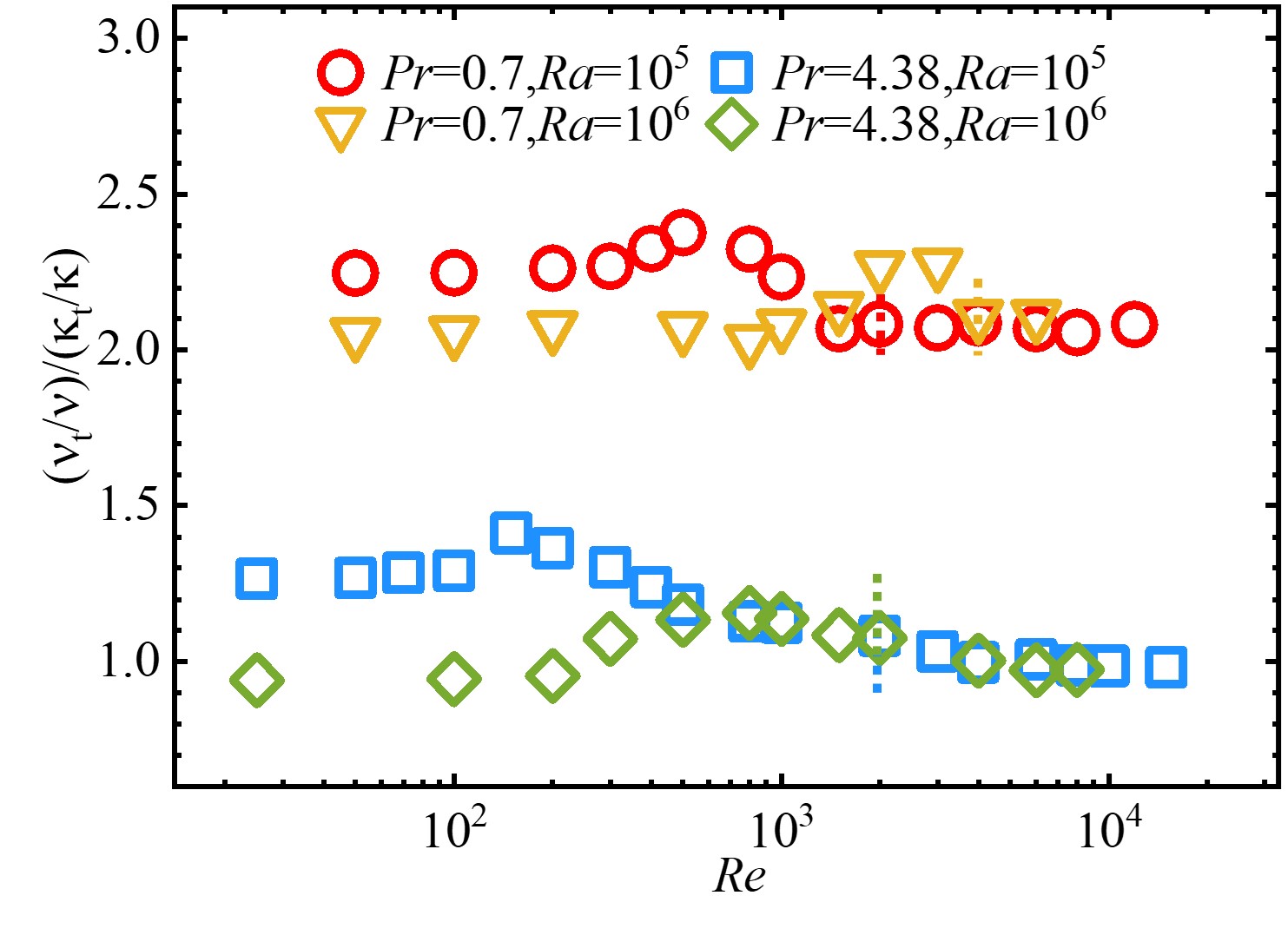}
 \caption{  \label{fig3} 
  The ratio of  $\nu_{t}/\nu$  over $\kappa_{t}/\kappa$ 
  as a function of $Re$. 
  Symbols are the same as those in {Fig.} \ref{fig2}.
  Vertical dotted lines denote the second transition $Re^*_2$.
 }
\end{figure}

Further analyses are performed regarding the similar properties
of heat and momentum transport
for high Reynolds number. 
In {Fig.} \ref{fig3}, we show
the ratio of the dimensionless effective viscosity 
$\nu_{t}/\nu$ to the diffusivity $\kappa_{t}/\kappa$ 
as a function of $Re$. 
The dimensionless effective diffusivity  
is defined as
${\kappa_{t}}/{\kappa}=aNu$,
where $a=2d^2/(ln(r_2/r_1)(r_2^2-r_1^2))$ 
is the  geometry factor (Appendix B). 
It is found that,
the ratio remains approximately 
a constant close to unity irrespective of $Ra$ for $Pr=4.38$
when $ Re\textgreater Re_2^*$. 
The ratio appears larger for a lower $Pr$.

For each set of $Ra$ and $Pr$,
we have seen from above that 
the Reynolds-number dependences of heat 
and angular momentum transport 
exhibit a similar trend, 
and the transitions of different transport regimes 
take place at same critical values
of $Re_1^*$ and $Re_2^*$. 
These results suggest that,
there is a unified mechanism that 
governs the processes of both heat and angular momentum transport 
in the present system.
In the following, 
we will gain some insights into the transport properties 
by analysing the flow morphology and structures 
in different flow regimes.

\subsection{\label{sec:level2} Flow morphology}
\noindent 

\begin{figure} [htbp]
 \centering
 \includegraphics[width=4.9in]{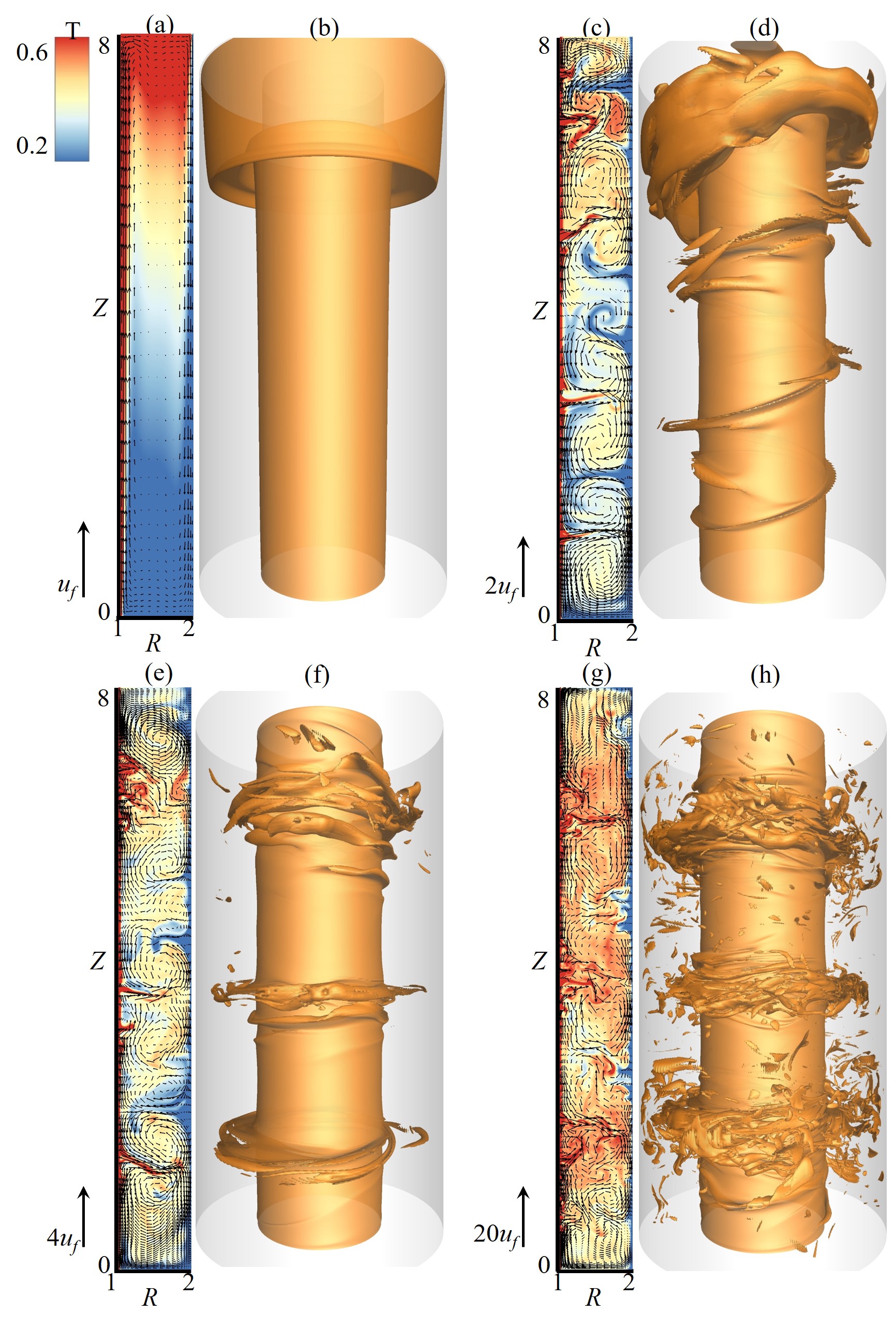}
 \caption{\label{fig4} 
  Two-dimensional  distributions
  of temperature and velocities 
  in the meridional plane (a,c,e,g) and
  three-dimensional temperature iso-surfaces ($T=0.52$)
  in the whole domain (b,d,f,h) for
  $Re$= 0 (a,b), 1000 (c,d), 2000 (e,f) and 4000 (g,h) 
  with $Ra= 10^{6}$ and $Pr=4.38$.
  Vertical arrow on the left side of each panel 
  denotes the velocity scale of free-fall velocity  
  $u_f=\sqrt{\beta g \Delta d}$.  
 }
\end{figure}

To illustrate the evolution of flow structures,
we show 
 two- and three-dimensional temperature
fields  in {Fig.} \ref{fig4}.
Without rotations 
($Re=0$ as seen in {Figs.} \ref{fig4}(a) and (b)), 
the convective structure is axisymmetric,
with a large-scale meridional circulation 
carrying ascending hot flows along the inner cylinder 
and descending cold flows along the outer cylinder.
These are the typical temperature distributions in vertical convection
\cite{Shishkina16PRE, XinPoF_vertical, lau2013large_vertical}. 
For $Re\textless Re_1^*$,
since the rotation is too weak to
 affect  the circulation,
 we thus see that the $Nu$ and $\nu_{t}/\nu$ 
 remain at their non-rotating values
  shown in {Figs.} \ref{fig1} and \ref{fig2}.  
When the rotation effect becomes comparable to the buoyancy
for $Re \ge Re^*_1$,
the flow undergoes the first transition 
from the meridional circulation
to spiral vortices 
(see {Figs.} \ref{fig4}(c) and (d)),
which starts to 
enhance the radial transport of heat and angular momentum.
In this rotation-affected regime, 
we find that 
both the flow morphology and the transport properties 
are dependent on both $Ra$ and $Pr$.
With further increase in  $Re$,
 rotation gradually dominates buoyancy,
and the spiral vortices are replaced 
by the toroidal TVs 
(see {Figs.} \ref{fig4}(e) and (f)).
In this TC-dominated regime 
$Nu(Re)$ and $\nu_{t}/\nu(Re)$ 
 collapse respectively onto
one curve that is independent
 of $Ra$ 
 ({Figs.} \ref{fig1}(a) and \ref{fig2}(a)). 
The spiral vortices exist in 
  an intermediate $Ri$-regime $0.089 \le Ri \le 2.54$
 depending on $Ra$ and $Pr$.
In a similar system \cite{Ball97}, 
 the onset of spiral flow is reported
in the range of $0.12 \textless Ri \textless 0.47$,
showing a reasonable consistency with the present results.
We note that
the convection states of co-rotating and counter-rotating vortices 
reported in Ref. \cite{Guillerm15} are not observed in the present study,
presumably due to the different geometry of fluid domain 
used in the present 
study.

\begin{table}
 \centering
 \caption{\label{tab3:Re_Ri}%
  A summary of the transitional values
  $Re_1^*$ ($Ri_1^*$) and
  $Re_2^*$ ($Ri_2^*$).}
 \begin{ruledtabular}
  \renewcommand\tabcolsep{5pt} 
  \begin{tabular}{lccc}
   &
   $Re_1^*$  ($Ri_1^*$) &
   $Re_2^*$  ($Ri_2^*$) &
   \\
   \hline 
   $Ra=10^5,Pr=0.7$& 
   500(0.57)& 
   2000(0.036)&  \\  
   $Ra=10^5,Pr=4.38$& 
   150(1.01)& 
   2000(0.006)&  \\  
   $Ra=10^6,Pr=0.7$& 
   1000(1.43)& 
   4000(0.089)&  \\  
   $Ra=10^6,Pr=4.38$& 
   300(2.54)& 
   2000(0.057)&  \\  
  \end{tabular}
 \end{ruledtabular}
\end{table}

We see that
with sufficiently large Reynolds number $Re\ge Re_2^*$,
the flow undergoes the second transition
to the turbulent TV flow
({Figs.} \ref{fig4}(g) and (h)),
 which is in good accord with previous studies 
 in adiabatic TC systems \cite{DONG2007Direct}
  and in TC systems 
  with a vertical temperature gradient applied \cite{Leng_2021}. 
In this flow regime, 
the turbulent transport of heat and momentum follows 
a  unifying power-law scaling 
as shown in {Figs.} \ref{fig1}(a) and \ref{fig2}(a). 
In contrast to the flow  fields with a lower 
$Re$ $(Re\textless Re_2^*)$,
we can see that the temperature field with $Re\ge Re_2^*$
 consists of rich small-scale structures,
with hot fluids
pumped more efficiently into the bulk flow 
from the inner cylinder,
resulting in  a higher transport efficiency.
Note that 
for $Ra=10^6$ and $Pr=0.7$
the vigorous buoyancy-driven turbulence
largely postpones the second transition $Re^*_2$.
We thus find that 
the spiral vortices
can  persist for $Re\textgreater2000$,
and then turn into the turbulent TVs directly
when $Re\textgreater4000$.

\subsection{\label{sec:level2} Mutual coherent structures for heat and angular 
momentum transport}

\begin{figure} [htbp]
 \centering
 \includegraphics[width=6.0in]{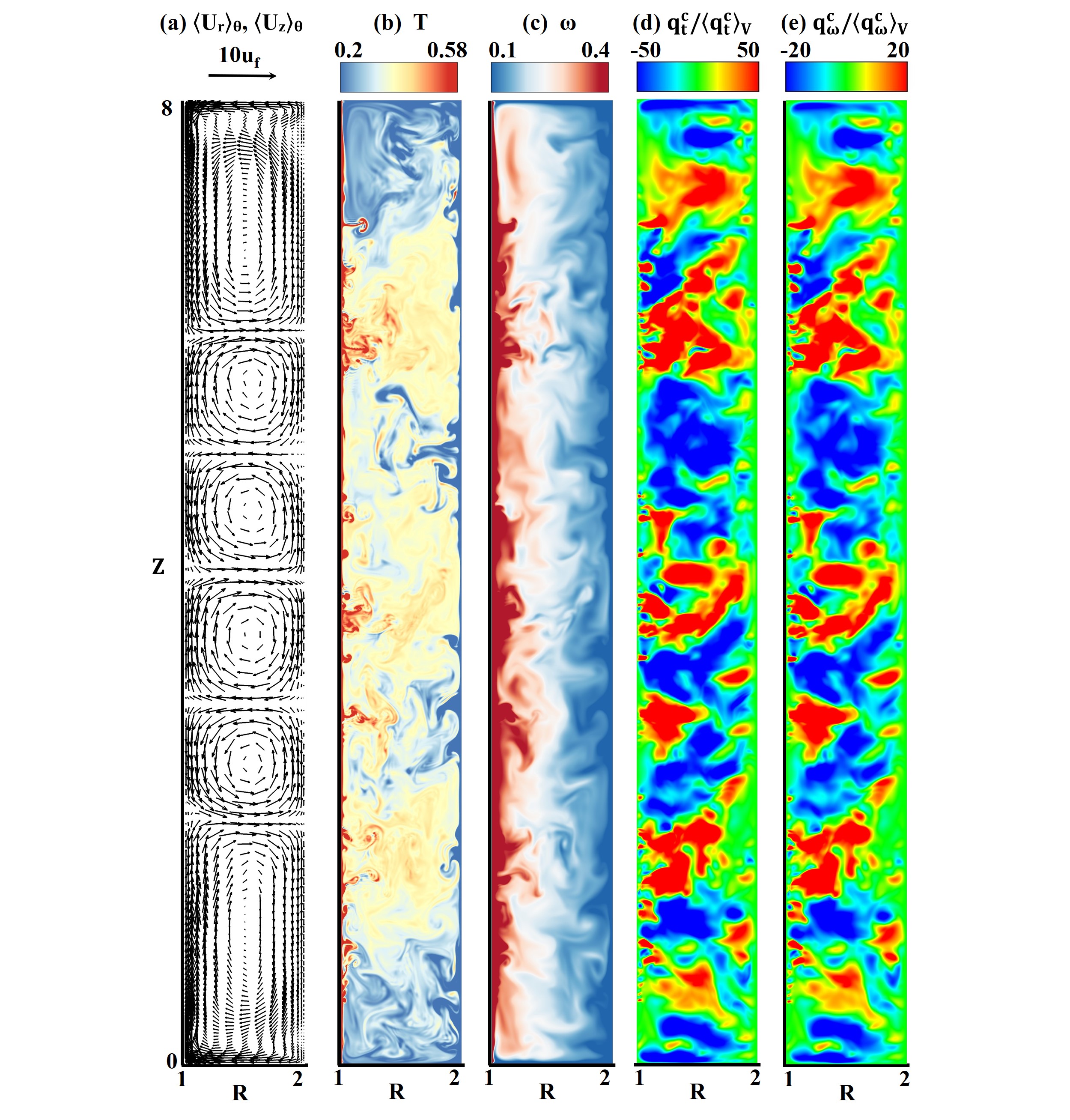}
 \caption{  \label{fig5} 
  Distributions of 
  time-averaged velocity field (a),
  instantaneous distributions of fluid
  temperature $T$ (b), 
  angular velocity $\omega$ (c), 
  the convective heat flux density 
  $q^c_t/ \left \langle q^c_t \right 
  \rangle_V$ (d)
  and the convective angular velocity flux density
  $q^c_\omega / \left \langle q^c_\omega \right \rangle_V$ (e) 
  in the same meridional plane for $Re=10000$, $Ra= 10^{5}$ and $Pr=4.38$.
 The arrow on the top of panel (a) denotes the 
  velocity scale of free-fall velocity $u_f$. 
  For comparisons, panels (d) and  (e) are plotted using the 
  same coloration.
 }
\end{figure}

In this section, 
we demonstrate that 
mutual coherent structures 
in forms of turbulent TVs exist,
 through which
 both heat and angular momentum are transported 
efficiently in the high-$Re$ flow regime.
To verify this viewpoint,
we present in {Fig.} \ref{fig5} 
the flow fields and
the spatial distributions of heat and angular velocity fluxes, 
respectively. 
In {Fig.} \ref{fig5}(a), 
three pairs of TVs characterize 
the time-averaged  velocity field.
The instantaneous temperature field shown in
{Fig.} \ref{fig5}(b),
however, is dominated  by  turbulent fluctuations.
TVs can be recognized roughly as 
the hot (cold) plumes which are emanating from
 the inner (outer) cylinder 
towards the bulk flow.
In the instantaneous fields of $\omega$ 
({Fig.} \ref{fig5}(c)), 
we observe large-scale coherent structures, 
while TVs 
are hard to be identified.  
In {Figs.} \ref{fig5}(d) and (e) 
we further show the spatial distributions of
convective flux densities of 
heat ($q^c_t/\left\langle q^c_t\right \rangle_{V}$) and 
angular velocity 
($q^c_{\omega}/\left\langle q^c_{\omega}\right \rangle_{V} $),
 normalized by their averaged values 
(see definitions of  flux densities  in Appendix B). 
We can see that
both the instantaneous flux densities exhibit 
a similar spatial distribution. 
Therefore, 
we conjecture  that
the turbulent heat and angular momentum transport 
are archived through mutual coherent structures
in the high-$Re$ regime.

To gain more insight into the mutual coherent structures 
for heat and angular momentum transport, 
we present data analysis of the spatial distribution 
of the local flux densities. 
We denote each
spatial position of the fluid domain studied as
$P(r, \theta, z)$.
Following the strategy used in Refs. \cite{Olga_07,Huang_PRl},
we identify the pronounced structures of 
efficient turbulent transport, 
determining the spatial regions 
where the flux densities ($q^c_t$, $q^c_{\omega}$) 
are greater than their averaged values
($\left \langle q^c_t \right \rangle_{V}$,
 $\left \langle q^c_{\omega} \right 
\rangle_{V}$).  
Thus, for heat transport
 we define
$(i)$  hot plumes  $P_{t,hot}(r, \theta, z)$ 
 where $q^c_t(r, \theta, z) \ge C\cdot 
 \left \langle q^c_t \right \rangle_{V}$ and
$(ii)$  cold plumes  $P_{t,cold}(r, \theta, z)$ 
 where $q^c_t(r, \theta, z) \le - C\cdot
 \left \langle q^c_t \right \rangle_{V}$.
Similarly, for  angular velocity transport,
 we define
$(iii)$  ``hot'' (positive) plumes 
$P_{{\omega},hot}(r, \theta, z)$ 
where $q^c_{\omega}(r, \theta, z) \ge C\cdot
 \left \langle q^c_{\omega} \right 
\rangle_{V}$ and
$(iv)$ ``cold'' (negative) plumes  
$P_{{\omega},cold}(r, \theta, z)$ 
where $q^c_{\omega}(r, \theta, z) \le - C\cdot
\left \langle q^c_{\omega} \right 
\rangle_{V}$.
The first subscript ($t$, $\omega$) 
denotes heat and angular velocity, 
and the second subscript ($hot$, $cold$) 
denotes the hot and
cold plumes  respectively. 
The factor $C$ is an empirical parameter chosen to be
in the range of $1\le C\le 40$ in this study.
The mutual coherent structures are then defined
as the overlapping volume of 
$P_{{t}}(r, \theta, z)$ and $P_{{\omega}}(r, \theta, z)$
as follows,
$(v)$ the mutual hot plumes 
$P_{t,\omega,hot}(r, \theta, z)$ where 
$q^c_t(r, \theta, z) \ge C\cdot \left \langle q^c_t \right \rangle_{V}$ 
and
$q^c_{\omega}(r, \theta, z) \ge C\cdot
\left \langle q^c_{\omega} \right 
\rangle_{V}$,
and 
$(vi)$ the mutual cold plumes 
$P_{t,\omega,cold}(r, \theta, z)$ where
$q^c_t(r, \theta, z) \le - C\cdot
\left \langle q^c_t \right \rangle_{V}$
and
$q^c_{\omega}(r, \theta, z) \le - C\cdot
\left \langle q^c_{\omega} \right 
\rangle_{V}$. 
Through time- and volume-averaging,  
we obtain the mean volumes of 
the hot  plumes
\begin{equation} 
\begin{aligned}
V_{t,hot} =(V_0\tau_0)^{-1}\int_{V,\tau}P_{t,hot}dVd\tau,\quad
V_{\omega,hot} =(V_0\tau_0)^{-1}\int_{V,\tau}P_{\omega,hot}dVd\tau,
\end{aligned}
\end{equation}
the cold plumes 
\begin{equation} 
 \begin{aligned}
V_{t,cold}=(V_0\tau_0)^{-1}\int_{V,\tau}P_{t,cold}dVd\tau,\quad
V_{\omega,cold}=(V_0\tau_0)^{-1}\int_{V,\tau}P_{\omega,cold}dVd\tau,
\end{aligned}
\end{equation}
and the mutual plumes 
\begin{equation} 
\begin{aligned}
V_{t,\omega,hot}=(V_0\tau_0)^{-1}\int_{V,\tau}P_{t,\omega,hot}dVd\tau,\quad
V_{t,\omega,cold}=(V_0\tau_0)^{-1}\int_{V,\tau}P_{t,\omega,cold}dVd\tau,
\end{aligned}
\end{equation}
where $V_0=\int_{V}PdV$ and $\tau_0$ denote the whole volume 
and time period.

\begin{figure} [htbp]
 \centering
 \includegraphics[width=6.4in]{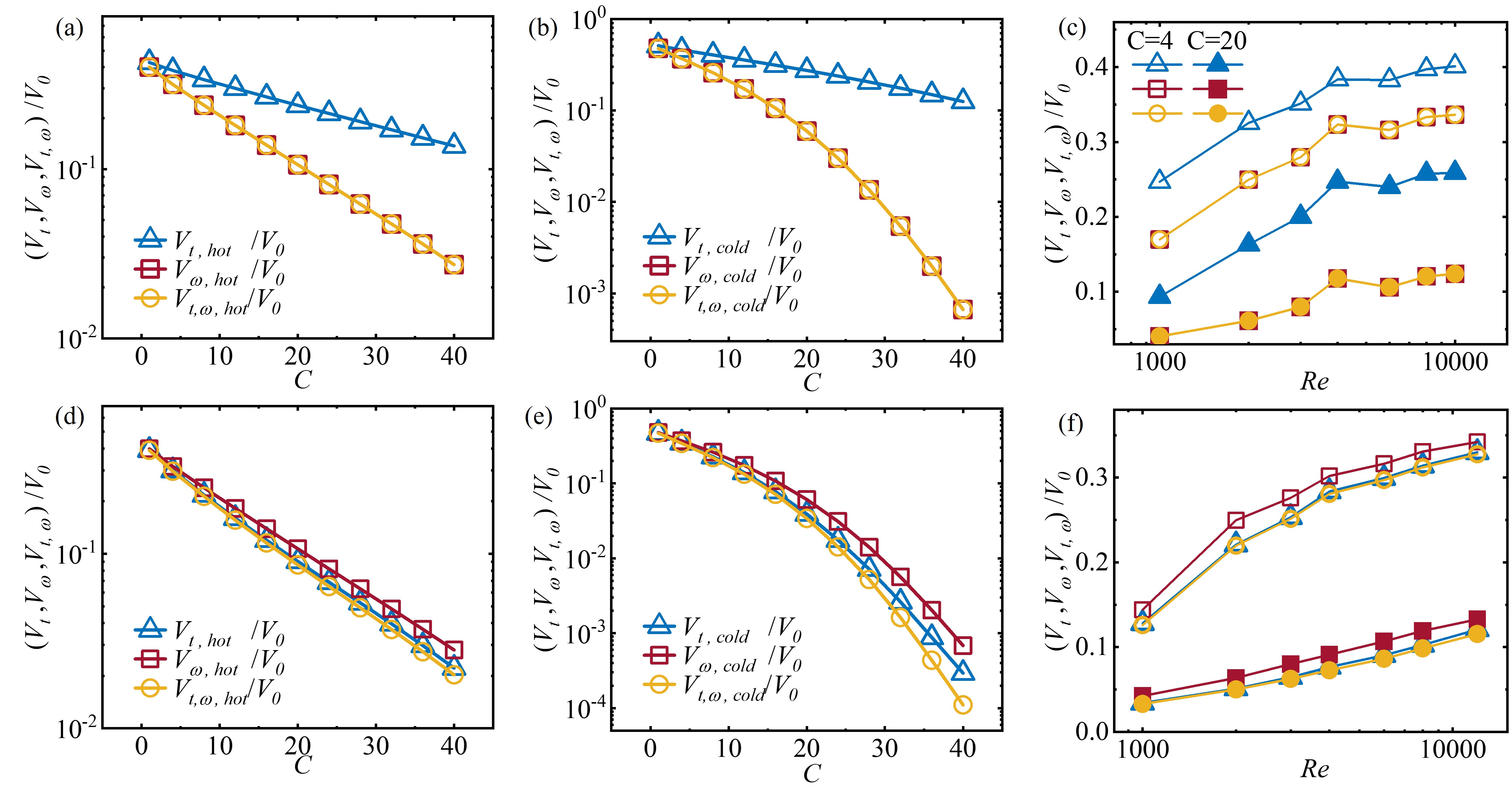}
 \caption{  \label{fig6} 
  Volume ratios of 
   the hot (positive) (a,d) and 
  cold (negative)  (b,e) plumes 
  for heat,  angular velocity and the mutual parts
  as functions of coefficient $C$ 
  for $Re=6000$.
  (c,f) Volume ratios of the hot (positive)  plumes
  as functions of $Re$ for various $C$. 
   For all panels we use $Ra=10^5$.
  Results in panels (a,b,c) are for $Pr=4.38$
   and (d,e,f) are for $Pr=0.7$.
 }
\end{figure}

The volume ratios 
$V_{t}/V_0$,
$V_{\omega}/V_0$ and
$V_{t,\omega}/V_0$ for $Pr=4.38$
 are plotted as functions of $C$ in
  {Figs.} \ref{fig6}(a) and (b). 
 As shown in {Fig.} \ref{fig6}(a),
 these  ratios for hot plumes are about 0.42 for $C=1$,
 and decrease as $C$ increases.
 Interestingly, 
 data of the ratios
 $V_{\omega,hot}/V_0$
 and  $V_{t,\omega,hot}/V_0$ collapse, 
 both decreasing more rapidly than $V_{t,hot}/V_0$.
 A similar trend is
 shown in {Fig.} \ref{fig6}(b)
  for the volume ratios of cold plumes.
 We see that the ratios of  thermal plumes 
 $V_{t,hot}/V_0$ and $V_{t,cold}/V_0$
 are always greater than 
 the angular-velocity and the mutual ones,
 indicating a broader distribution of thermal structures.  
In {Fig.} \ref{fig6}(c),            
the volume ratios of hot (positive) plumes
 are plotted as functions of $Re$ for two values of $C$. 
We see that 
with increasing $Re$ the ratios
$V_{t,hot}/V_0$,
$V_{\omega,hot}/V_0$ and
$V_{t,\omega,hot}/V_0$
first increase and then become 
independent of $Re$ when $Re\ge4000$.
 
 For the flows with low $Pr=0.7$   
({Figs.} \ref{fig6}(d), (e) and (f)),
the data show almost  similar trends 
when the parameters $C$ and $Re$ change.
 We see that 
 the volume ratios become greater when $Re$ increases, 
 or when $C$ decreases.
 However, here we find that $V_{\omega}/V_0$
 becomes slightly greater than $V_t/V_0$ and $V_{t,\omega}/V_0$
for low $Pr$.
 We attribute these to the $Pr$-dependence of the flow properties,
 since heat is more likely to accumulate within 
 the turbulent coherent structures for high $Pr=4.38$,
but becomes  easier to diffuse for  low $Pr=0.7$. 
 Results in {Fig.} \ref{fig6} imply that 
 in the high-$Re$ regime heat and angular momentum 
 are transferred mainly through highly similar 
 coherent structures. 
 The flow regions of large angular momentum fluxes are nested 
 within the regions of large heat fluxes for high $Pr=4.38$,
 and vice versa for low $Pr=0.7$. 
 We suggest that 
 it is the similarities of the turbulent structures, 
 which deliver efficiently both 
 the heat and angular momentum transport,
 give rise to the same scaling properties of 
 $Nu$ and $\nu_{t}/\nu$ observed 
 in {Figs.} \ref{fig1}, \ref{fig2} and \ref{fig3}.

\section{\label{sec:level1} Concluding remarks}

We investigate numerically
 the heat and angular momentum transport processes
in the turbulent Taylor-Couette flows
which are subjected to a radial temperature gradient.
A large range of Reynolds number is considered,
extending the present study of the heat transport to 
the unexplored regime of turbulent TVs.

We find that 
the flows undergo a first transition at $Re^*_1$ 
from the convection-dominated state
in the form of a large-scale meridional circulation 
to the transitional regime typified by spiral vortices.
After this transition 
we observe enhanced transport of  heat and 
angular momentum, 
since rotations start to influence the flow structures. 
With increasing $Re$, 
the flow turns into the TC-dominated regime 
where the heat and angular momentum transport 
 become independent of $Ra$.
Eventually, 
the turbulent TVs start to dominate 
the turbulent transport processes
at the second transition $Re^*_2$,
after which the heat and angular momentum transport
are dictated by power-law scalings, 
i.e.,
$Nu\sim Re^{0.619\pm0.015}$ for $Pr=4.38$,
$Nu\sim Re^{0.590\pm0.025}$ for $Pr=0.7$ and 
$\nu_{t}/\nu\sim Re^{0.588\pm0.036}$ for both $Pr$.
Our results also show that 
the transitional values $Re^*_1$ and $Re^*_2$ 
depend on both $Ra$ and $Pr$.

A striking finding is the analogy between the
radial transport of heat and angular momentum.
Besides their similar scaling exponents,
our data show that 
the effective viscosity ($\nu _t/\nu$) and
 diffusivity ($\kappa_t/\kappa$) 
have almost the same efficiency for 
$ Re \textgreater 2000$.
The similar properties of both types of transport 
 are found to persist 
in the turbulent TV regime,
which was attributed to  
 the mutual structures 
through which heat and momentum are efficiently transported.
Further analysis  
shows that 
the structures for high-efficiency angular momentum transport 
 are nested inside the thermal ones for high $Pr=4.38$, 
 or vice versa for low  $Pr=0.7$.
We note that
the analogy between heat and momentum transport in rotating flow 
has been interpreted  through
the one-dimensional simplified model by Bradshaw \cite{bradshaw1969analogy}. 
To further connect the present results with Bradshaw’s analogy 
is an intriguing subject for future studies.
In a  TC system
where an axial destabilized temperature gradient is applied,
it has been reported that,
the axial heat transport  has the same scaling 
as the radial angular momentum transport 
in the turbulent TV regime \cite{Leng_2021}.
Hence, we suggest that 
it is the structures in forms of turbulent TVs that
 provide the TC systems with the equal transport efficiencies
in both the radial and axial directions.

The ultimate regime of TC flows 
\cite{Huisman2012,zhu2018wall}
 sets in 
at a much larger Reynolds number 
($Re\textgreater6\times10^4$ for $\eta=0.5$
\cite{Merbold2013,ostilla2014exploring,Veen2016Taylor}) than
 the parameters considered in the present study.
Whether a similar scaling of  heat and angular momentum 
transport exists
at higher $Re$  and even in the ultimate regime,  
remains a challenging problem for future studies.

\begin{acknowledgments}
 This work was supported by the Natural Science Foundation of China under grant
 nos. 11902224, 11772235 and the China Postdoctoral Science 
 Foundation (No.2019M651572).
\end{acknowledgments}

\appendix
\section{Numerical details}

\subsection{\label{app:Results} Grid sensitivity studies and main results}

The results of the grid sensitivity studies are listed in {Table} 
\ref{tab1:Res}.
It is shown that,  
 results of $Nu$ and  $Re_{\tau}$ 
 show a good convergence as  resolutions increase.
The main results are listed in  {Tables} 
\ref{tab2:results_4.38}  and  \ref{tab3:results_0.7}.
For all our runs, the smallest mean scales are respectively determined 
by the mean Kolmogrov scale 
$\left \langle \lambda_k \right \rangle_{V} = 
(\nu^3/\left\langle {\epsilon_\nu}\right\rangle_{V} )^{1/4}$ 
for $Pr=0.7$,
and  the mean Batchelor scale 
$\left\langle  \lambda_b\right\rangle_{V} 
=(\kappa^2\nu/\left\langle {\epsilon_\nu}\right\rangle_{V} )^{1/4}$
for $Pr=4.38$, where $\left\langle {\epsilon_\nu}\right\rangle_{V}$ 
is the volume- and time-averaged turbulent kinetic energy dissipation rate 
\cite{batchelor1959small, pope2000turbulent, 
 stevens2010radial, scheel2013resolving}. 
At high Reynolds number, the flow enters into the shear-dominated regime, 
that $\left\langle \lambda_b\right\rangle_{V}$ decreases rapidly with the 
increasing kinetic energy dissipation. 
Thus, the ratios of the greatest grid spacing $L_{max}$
\cite{emran2008fine}
to $\left\langle  \lambda_k\right\rangle_{V}$  and 
$\left\langle  \lambda_b\right\rangle_{V} $
\cite{grotzbach1983spatial}
do not exceed 2.7 and 4.5 respectively in this study.
Meanwhile, the minimal radial grid spacing in wall units is always less than 
unit. 
Besides this, 
the relative error measurement $\sigma _{ \epsilon _T}$ 
is employed to check the deviation
of the exact balance between the thermal dissipation and 
the  global  heat transfer
\cite{shraiman1990heat,stevens2010radial,scheel2013resolving}.
To reduce the computational requirements for high-$Re$ flows,  
the azimuthal computational extents $L_{\theta}$ are reduced to a quarter of 
the cylinder ($0.5\pi$) for $Re \ge 4000$. 
This strategy has been proven to be effective 
\cite{brauckmann_eckhardt_2013, Ostilla_2015_PoF}.
And as shown in   {Tables} 
\ref{tab2:results_4.38}  and  \ref{tab3:results_0.7}, 
the simulations for $L_{\theta}=2\pi$ and $0.5\pi$ 
are both performed 
in the range of $1000 \le Re \le 4000$, 
and the results suggest that 
the shortened extents do not change the main results.

In {Tables} \ref{tab1:Res},
 \ref{tab2:results_4.38} and \ref{tab3:results_0.7},
$N_{\theta}{\times}{N_z}{\times}N_r$ denote the resolutions 
in three directions, and
$L_{\theta}$ is the azimuthal computational extent;
$\sigma _{ \epsilon _T}$ is the relative error measured by $Nu$ 
and the thermal dissipation rate; 
$\frac{L_{max}} {\left \langle \lambda_k \right \rangle_{V}, 
 \left \langle\lambda_b \right \rangle_{V}}$ are the 
maximal grid 
spacings compared with the Kolmogrov and 
Batchelor scales;
$r^+_{min}, r^+_{max}$ are the minimal and maximal 
grid sizes in wall units.

\begin{figure}[ht]
 \includegraphics[width=5in]{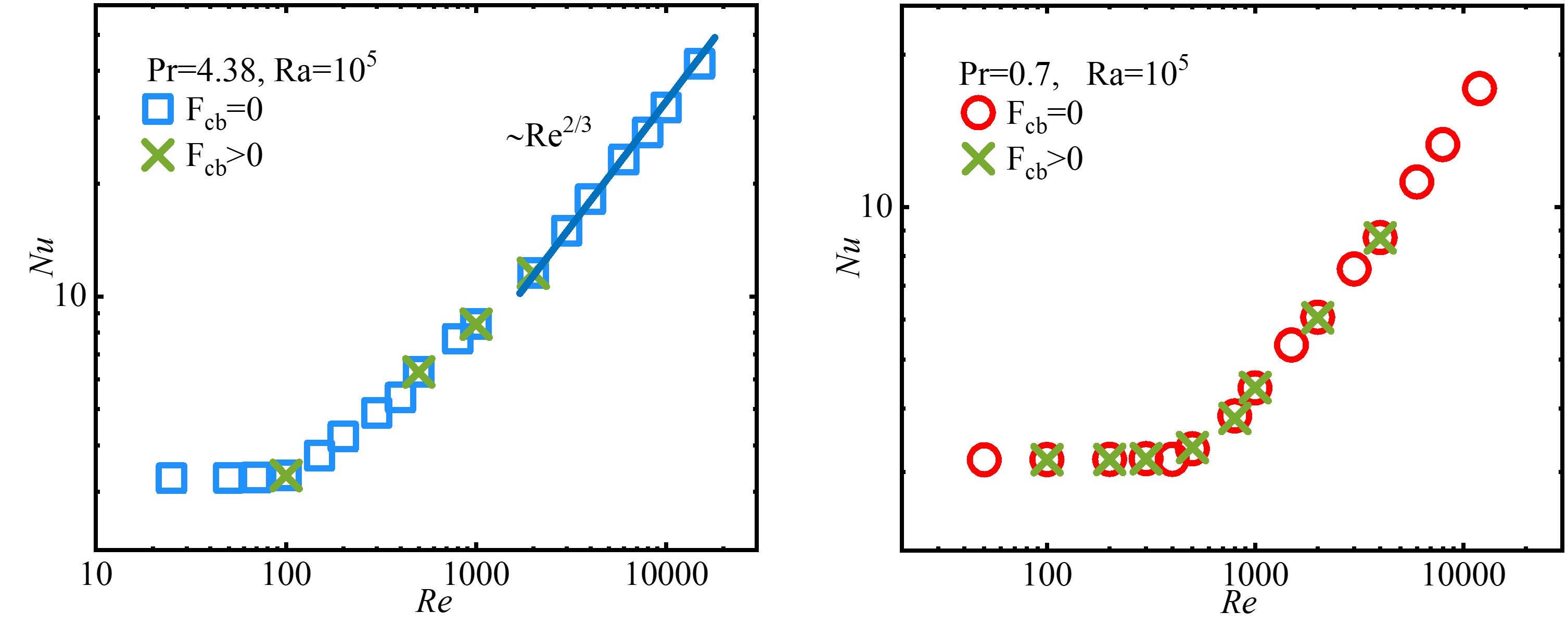}
 \caption{\label{fig7}  
  Comparison of 
  $Nu$ as functions of $Re$ 
  when the  centrifugal buoyancy is included 
  ($F_{cb}\textgreater0$) and 
  excluded ($F_{cb}=0$). 
  Results for  $Pr=4.38$ (a), $Pr=0.7$ (b) and $Ra=10^5$.
  Solid line indicates the scaling law obtained from experiments
  \cite{Fujio64}. 
 }
\end{figure}

\subsection{\label{app:Validation} Discussions of the centrifugal buoyancy 
effects}

In our system, 
the centrifugal buoyancy 
$F_{cb}=-(\beta \Delta)({U_{\theta}^2}/{R})T{\mathit{e_r}}$
\cite{Yoshikawa13, Guillerm15}
is present
because of the azimuthal motion of the fluid.   
Here we perform the additional simulations for 
the experimental conditions 
with $\beta \approx 0.004K^{-1}$ and $\Delta = 10K$ for air
and with $\beta \approx 0.00038K^{-1}$ and $\Delta = 10K$ for water.
The results shown in {Fig.} \ref{fig7} indicate that, 
including the centrifugal buoyancy does not 
change the results of heat transport.
Therefore, in this paper, 
the effect of centrifugal force is neglected. 

\subsection{\label{app:Comparision} Comparison with experimental results}

To validate our results for high-$Re$ regime, 
we compare the heat-transport data with 
the previous experimental results in
{Fig.} \ref{fig6}(a). 
Our results are consistent with the power-law scaling obtained in Ref.
\cite{Fujio64} for $2000\le Re \textless10000$. 
But for high $Re \textgreater 10000$, 
it is found that $Nu$ tends to deviate from the scaling law.
We argue that the difference results from their fitting errors, 
since this scaling exponent 
is already same to the  well-accept value ($\textgreater2/3$) 
for ultimate turbulent regime \cite{grossmann2016high}.

\begin{table}
 \centering
 \caption{\label{tab1:Res}%
  A summary of the grid sensitivities study for $Ra=10^5$, $Re=8000$ and 
  $Pr=0.7$.}
 \begin{ruledtabular}
  \renewcommand\tabcolsep{5pt} 
  \begin{tabular}{lccccc}
   $N_{\theta}(L_{\theta}) {\times}N_z {\times}N_r$ &
   $Nu$ &
   $Re_{\tau}$ &
   $\sigma_{\epsilon_T}$ &
   $L_{max}/(\left \langle \lambda_k \right \rangle_{V},
   \left \langle \lambda_b\right \rangle_{V})$ &
   \\
   \hline  
   $128(0.5\pi)\times1025\times225$& 
   13.54 & 
   232.95& 
   0.02 & 
   2.45,  2.05  &  \\
   $192(0.5\pi)\times1537\times225$& 
   13.29 & 
   229.24 & 
   0.01 & 
   1.86,  1.56  &  \\
   $256(0.5\pi)\times2049\times225$& 
   13.20 & 
   229.53 & 
   0.01 & 
   1.51,  1.26  &  \\   
  \end{tabular}
 \end{ruledtabular}
\end{table}

\begin{table}
 \centering
 \caption{\label{tab2:results_4.38}%
  A summary of the main results for $Pr=4.38$.}
 \begin{ruledtabular}
  \renewcommand\tabcolsep{5pt} 
  \begin{tabular}{lccccccccccc}
   $Re$ &
   $Pr$ &
   $Ra$ &
   $N_{\theta}(L_{\theta}) {\times}N_z {\times}N_r$ &
   $Nu$ &
   $Re_{\tau}$ &
   $\sigma_{\epsilon_T}$ &
   $L_{max}/(\left \langle \lambda_k \right \rangle_{V},
   \left \langle \lambda_b\right \rangle_{V})$ &
   $r^+_{min},r^+_{max}$ &
   \\
   \hline    
   0&  4.38& $10^5$& $128(2\pi)\times257\times33$& 3.27 & $-$&  
   0.00&  
   0.90, 1.89&  $-$, $-$&  \\   
   25&  4.38& $10^5$& $128(2\pi)\times257\times33$& 3.27 & 4.99&  
   0.00&  
   0.91, 1.91&  0.07, 0.59&  \\
   50&  4.38& $10^5$& $128(2\pi)\times257\times33$& 3.27&  7.06&  0.00 
   & 
   0.94,  1.96&  0.10,  0.84&  \\ 
   70&  4.38& $10^5$& $128(2\pi)\times257\times33$& 3.28&  8.42& 
   0.00&  
   0.97,  2.02 & 0.12,  1.00&  \\ 
   100& 4.38& $10^5$& $128(2\pi)\times257\times33$& 3.32&  10.16&  
   0.00 & 
   1.02,  2.13 & 0.14,  1.20&  \\ 
   150& 4.38& $10^5$& $128(2\pi)\times257\times33$& 3.75&  13.83&  
   0.01 & 
   1.14,  2.39 & 0.19,  1.64&  \\ 
   200& 4.38& $10^5$& $128(2\pi)\times321\times41$& 4.22&  16.64&  
   0.01&  
   1.27,  2.66 & 0.23,  1.72&  \\ 
   300& 4.38& $10^5$& $128(2\pi)\times321\times41$& 4.88&  21.41&  
   0.01&  
   1.33,  2.79 & 0.23,  1.77&  \\ 
   400& 4.38& $10^5$& $128(2\pi)\times321\times41$& 5.37&  25.29&  
   0.02 & 
   1.54,  3.22 & 0.27,  2.40&  \\ 
   500& 4.38& $10^5$& $128(2\pi)\times321\times41$& 6.28&  29.89&  
   0.02 & 
   1.77,  3.69 & 0.32,  2.83&  \\ 
   800& 4.38& $10^5$& $128(2\pi)\times385\times49$& 7.70&  40.88&  
   0.02 & 
   2.05,  4.29 & 0.36,  3.23&  \\

   1000& 4.38& $10^5$& $192(2\pi)\times513\times65$&  8.43 & 47.74&  
   0.02&  1.76,  3.69&  0.23,  3.09&  \\ 
   1000& 4.38& $10^5$& $64(0.5\pi)\times513\times65$&  8.47 & 47.89&  
   0.02&  1.60,  3.36&  0.23,  3.10&  \\ 
   2000& 4.38& $10^5$& $256(2\pi)\times721\times91$ & 11.63 & 77.46&  
   0.03 & 1.93,  4.04&  0.26,  3.57&  \\ 
   2000& 4.38& $10^5$& $96(0.5\pi)\times721\times91$&  11.74&  77.65& 
   0.03&  1.69,  3.54&  0.26,  3.58&  \\ 
   3000& 4.38& $10^5$& $128(0.5\pi)\times897\times129$ & 14.98 & 
   105.63 & 
   0.04 & 1.64,  3.42&  0.25,  3.42&  \\ 
   4000& 4.38& $10^5$& $128(0.5\pi)\times1025\times161$& 18.16 & 
   131.90 & 
   0.04 & 1.74,  3.65&  0.25,  3.42&  \\ 
   6000& 4.38& $10^5$& $192(0.5\pi)\times1281\times193$& 23.19 & 
   183.50 & 
   0.04 & 1.73,  3.62&  0.29,  3.96&  \\ 
   8000& 4.38& $10^5$& $256(0.5\pi)\times1537\times225$& 27.40&  
   228.16&  
   0.03 & 1.69,  3.53&  0.30,  4.20&  \\ 
   10000& 4.38& $10^5$& $256(0.5\pi)\times1793\times257$& 31.90 & 
   275.04 & 
   0.04 & 1.77,  3.71&  0.32,  4.46&  \\ 
   15000& 4.38& $10^5$& $256(0.5\pi)\times2001\times301$& 41.79 & 
   384.76&  
   0.04 & 2.11,  4.43&  0.38,  5.32&  \\

   0 & 4.38& $10^6$& $256(2\pi)\times513\times65$& 5.75&  $-$ & 0.00 
   & 
   0.93,  1.95 & $-$,  $-$&  \\ 
   25 & 4.38& $10^6$& $256(2\pi)\times513\times65$& 5.75&  5.70 & 0.00 
   & 
   0.93,  1.95 & 0.04,  0.34&  \\ 
   100& 4.38& $10^6$& $256(2\pi)\times513\times65$& 5.75 & 11.42&  
   0.00 & 
   0.94,  1.97 & 0.07,  0.68&  \\ 
   200& 4.38& $10^6$& $256(2\pi)\times513\times65$& 5.76 & 16.25 & 
   0.00 & 
   0.97,  2.03 & 0.11,  0.96&  \\ 
   300& 4.38& $10^6$& $256(2\pi)\times513\times65$& 6.21 & 21.93 & 
   0.00 & 
   1.01,  2.11 & 0.14,  1.30&  \\ 
   500 & 4.38& $10^6$& $256(2\pi)\times513\times65$& 6.89 & 30.66 & 
   0.01 & 
   1.15,  2.42 & 0.20,  1.82&  \\ 
   800& 4.38& $10^6$& $256(2\pi)\times513\times65$& 7.92 & 41.98 & 
   0.02 & 
   1.46,  3.06 & 0.20,  2.72&  \\ 
   1000& 4.38& $10^6$& $256(2\pi)\times513\times65$& 8.67 & 48.69 & 
   0.03 & 
   1.64,  3.43 & 0.24,  3.15&  \\ 
   1500& 4.38& $10^6$& $256(2\pi)\times641\times81$& 10.37 & 63.72 & 
   0.03 & 
   1.77,  3.69 & 0.24,  3.25&  \\ 
   2000& 4.38& $10^6$& $384(2\pi)\times721\times91$& 11.75 & 77.97 & 
   0.02 & 
   1.69,  3.55 & 0.27,  3.59&  \\ 
   
   4000& 4.38& $10^6$& $128(0.5\pi)\times1025\times161$& 17.97 & 
   131.67 & 
   0.04 & 1.74,  3.65&  0.25,  3.42&  \\ 
   6000& 4.38& $10^6$& $192(0.5\pi)\times1281\times193$& 23.38 & 
   181.11 & 
   0.04 & 1.72,  3.60&  0.28,  3.90&  \\ 
   8000& 4.38& $10^6$& $256(0.5\pi)\times1537\times225$& 27.40&  
   228.16&  
   0.03 & 1.69,  3.53&  0.30,  4.20&  \\ 
  \end{tabular}
 \end{ruledtabular}
\end{table}

\begin{table}
 \centering
 \caption{\label{tab3:results_0.7}%
  A summary of the main results for $Pr=0.7$.}
 \begin{ruledtabular}
  \renewcommand\tabcolsep{5pt} 
  \begin{tabular}{lccccccccccc}
   $Re$ &
   $Pr$ &
   $Ra$ &
   $N_{\theta}(L_{\theta}) {\times}N_z {\times}N_r$ &
   $Nu$ &
   $Re_{\tau}$ &
   $\sigma_{\epsilon_T}$ &
   $L_{max}/(\left \langle \lambda_k \right \rangle_{V},
   \left \langle \lambda_b\right \rangle_{V})$ &
   $r^+_{min},r^+_{max}$ &
   \\
   \hline

   0&  0.7& $10^5$& $128(2\pi)\times257\times33$& 3.17 & $-$ & 0.00 
   & 
   2.36,  1.98 & $-$,  $-$ &  \\
   50&  0.7& $10^5$& $128(2\pi)\times257\times33$& 3.17 & 9.26 & 0.00 
   & 
   2.37,  1.99 & 0.10,  1.19 &  \\
   100& 0.7& $10^5$& $128(2\pi)\times257\times33$& 3.17 & 13.06 & 0.00 
   & 
   2.38,  1.99 & 0.14,  1.69 &  \\
   200& 0.7& $10^5$& $128(2\pi)\times321\times41$& 3.18 & 18.59 & 0.00 
   & 
   2.09,  1.75 & 0.15,  1.92 &  \\
   300& 0.7& $10^5$& $128(2\pi)\times321\times41$& 3.19 & 22.83 & 0.00 
   & 
   2.15,  1.80 & 0.18,  2.36 &  \\
   400& 0.7& $10^5$& $128(2\pi)\times321\times41$& 3.18 & 26.66 & 0.00 
   & 
   2.23,  1.86 & 0.22,  2.76 &  \\
   500& 0.7& $10^5$& $128(2\pi)\times321\times41$& 3.33 & 30.87 & 0.01 
   & 
   2.30,  1.92 & 0.25,  3.19 &  \\
   800& 0.7& $10^5$& $128(2\pi)\times385\times49$& 3.87 & 41.59 & 0.01 
   & 
   2.32,  1.94 & 0.28,  3.58 &  \\

   1000 & 0.7& $10^5$& $192(2\pi)\times513\times65$ & 4.40  &48.59   & 
   0.01 & 2.10,  1.76 & 0.23,  3.12 &  \\
   1000 & 0.7& $10^5$& $64(0.5\pi)\times513\times65$ & 4.38 & 48.19 & 
   0.01 & 1.65,  1.38 & 0.23,  3.09 &  \\
   1500& 0.7& $10^5$& $192(2\pi)\times641\times81$& 5.34 & 63.12 & 
   0.01 & 
   1.95,  1.63 & 0.24,  3.27&  \\ 
   2000& 0.7& $10^5$& $256(2\pi)\times721\times91$ & 6.06 & 77.91 & 
   0.01 & 2.14,  1.79 & 0.27,  3.59 &  \\
   2000& 0.7& $10^5$ &$96(0.5\pi)\times721\times91$ & 6.10 & 78.25 & 
   0.00 & 1.94,  1.63 & 0.27,  3.60 &  \\
   3000& 0.7& $10^5$& $96(0.5\pi)\times897\times113$ & 7.55 & 106.18 & 
   0.01 & 1.89,  1.58 & 0.29,  3.93 &  \\
   4000& 0.7& $10^5$& $256(2\pi)\times1025\times161$ & 8.70 & 132.14 & 
   0.01 & 2.20,  1.84 & 0.25,  3.43 &  \\
   4000& 0.7& $10^5$ &$128(0.5\pi)\times1025\times161$& 8.88 & 133.44 
   & 
   0.01 & 1.75,  1.47 & 0.25,  3.46 &  \\
   6000& 0.7& $10^5$& $128(0.5\pi)\times1281\times193$& 11.20 & 
   182.83& 
   0.01 & 1.98,  1.66 & 0.28,  3.95 &  \\
   8000& 0.7& $10^5$ &$192(0.5\pi)\times1537\times225$& 13.29 & 229.24 
   & 
   0.01 & 1.86,  1.56 & 0.31,  4.25 &  \\
   12000& 0.7& $10^5$ &$192(0.5\pi)\times1793\times257$& 17.13 & 
   320.74 & 
   0.02 & 2.20,  1.84 & 0.37,  5.20 &  \\

   0 & 0.7& $10^6$& $256(2\pi)\times513\times65$& 5.49&  $-$ & 0.00 
   & 
   2.41,  2.01 & $-$,  $-$&  \\
   50 & 0.7& $10^6$& $256(2\pi)\times513\times65$& 5.49&  11.64 & 0.00 
   & 
   2.41,  2.00 & 0.06,  0.75&  \\ 
   100& 0.7& $10^6$& $256(2\pi)\times513\times65$& 5.48 & 16.47&  
   0.00 & 
   2.41,  2.00 & 0.08,  0.89&  \\ 
   200& 0.7& $10^6$& $256(2\pi)\times513\times65$& 5.48 & 23.34 & 
   0.00 & 
   2.42,  2.03 & 0.11,  1.5&  \\    
   500 & 0.7& $10^6$& $256(2\pi)\times513\times65$& 5.48 & 37.44 & 
   0.00 & 
   2.45,  2.05 & 0.18,  2.40&  \\    
   800& 0.7& $10^6$& $256(2\pi)\times513\times65$& 5.52 & 46.68 & 
   0.00 & 
   2.48,  2.07 & 0.23,  3.02&  \\    
   1000& 0.7& $10^6$& $256(2\pi)\times513\times65$& 5.58 & 68.53 & 
   0.00 & 
   2.54,  2.13 & 0.26,  3.44&  \\    
   1500& 0.7& $10^6$& $256(2\pi)\times513\times65$& 6.11 & 68.60 & 
   0.01 & 
   2.66,  2.22 & 0.33,  4.44&  \\    
   2000& 0.7& $10^6$& $256(2\pi)\times721\times91$& 6.55 & 84.49 & 
   0.01 & 
   2.29,  1.92 & 0.29,  3.89&  \\ 
   
   3000& 0.7& $10^6$& $256(2\pi)\times721\times91$& 7.67 & 112.20 & 
   0.01 & 
   2.68,  2.24 & 0.38,  5.17&  \\ 
   
   4000& 0.7& $10^6$& $256(2\pi)\times1025\times161$& 8.84 & 
   133.96 & 
   0.02 & 1.89,  2.26&  0.25,  3.47&  \\

   6000& 0.7& $10^6$& $128(0.5\pi)\times1281\times193$& 11.12 & 
   184.46 & 
   0.02 & 1.73,  3.62&  0.29,  3.96&  \\

  \end{tabular}
 \end{ruledtabular}
\end{table}

\section{ Derivations of flux densities, effective viscosity and diffusivity   }

In our annular system, 
the ring surface increases as $r$ increases, 
leading to a decreasing flux density along the radial direction.
Hence, in this section  the  
heat flux density $q_t$ and
 the angular velocity flux density $q_\omega$ are defined respectively,
 in addition to the common definitions of the heat and angular velocity 
 currents.

\subsection{\label{app:Heat} Heat flux density $q_{t}$}

Here, we consider the dimensional fields of
velocity $\textbf{u}{(r,{\theta},z)}$ and 
temperature  $t{(r,{\theta},z)}$.
For a pure thermal conductive state between the concentric cylinders, 
the one-dimensional radial temperature distribution is 
$t(r)=c_1 lnr+c_2$, 
with $c_1=-{\Delta}/{ln ({r_2}/{r_1})}$ and 
$c_2=t_2-c_1ln(r_2)$.
From Fourier's law, 
the radial heat flux density (by thermal conduction) is 
$\mathit{q_t^{Lam}(r)}=
-\kappa {\partial_r t}=-\kappa {c_1}/{r}$,
which decreases along the radial direction owing to the 
enlarging ring surface. 
For the turbulent flow, the 
three-dimensional distributions of the heat flux 
density is defined as,
\begin{equation}
  \mathit{q_t{(r,{\theta},z)}}
  = {u_r}{(r,{\theta},z)} 
  (t{(r,{\theta},z)}-t_2) - 
  \kappa 
  \partial_r{(t{(r,{\theta},z)}-t_2)}, 
\end{equation}
where the temperature of the cold wall  $t_2$  
is used as the reference temperature  
as done in Ref. \cite{kang2015}.
The first term on the right-hand side corresponds to
the convective contribution  
$q^c_t={u_r}{(r,{\theta},z)} 
(t{(r,{\theta},z)}-t_2)$.
Thus the Nusselt number is defined as 
the ratio of the turbulent heat transport to
the thermal conduction
\begin{equation}
 Nu=
 \frac{
 \left \langle q_t \right \rangle_{V}
 }
 { \left \langle q_t^{Lam} \right \rangle_{V}}
 =a^{-1}
 ({{\mathit{RePr}} 
  \left \langle U_rT\right\rangle_{V} 
  - \partial _{r}\left \langle 
  {T}\right\rangle_{V}}),
\end{equation}
where $a=2d^2/(ln(r_2/r_1)(r_2^2-r_1^2))$
denotes the factor caused by the annular geometry.

\subsection{\label{app:Angular} Angular velocity flux density $q_{\omega}$}

In circular Couette flow (CCF), the flow is laminar and 
has purely azimuthal velocity $u_{\theta}=A r+{B}/{r}$,
with $A=({r_2 u_2-r_1 u_1})/({r_2^2-r_1^2})$ and
$B=({r_2^2 r_1 u_1-r_2 r_1^2 u_2})/({r_2^2-r_1^2})$. 
Thus, its angular velocity flux density could be written as
$\mathit{q_{\omega}^{Lam}(r)}=2 \nu  {B}/{r}$,
which also decreases along the radial direction as same as $q_t^{Lam}$. 
Taking into account the increasing area in the radial direction, 
after multiplication with $r$, the conventional formula 
of the angular velocity current (for CCF) is
$J_{\omega}^{Lam}=2\nu B$ \cite{eckhardt2007torque}.
In turbulent flows, 
according to the derivations from Ref. \cite{eckhardt2007torque},
the conventional angular velocity current is expressed as
\begin{equation}
 \begin{aligned}
  \mathit{\left \langle J_{\omega}{(r,{\theta},z)}\right \rangle_{\theta z}} 
  =r^2 \left \langle
  {u_r}{(r,{\theta},z)} 
  {u_{\theta}}{(r,{\theta},z)}
  \right \rangle_{\theta z} - 
  \nu r^3 
  \partial_r
  \left(
  \left \langle u_{\theta}{(r,{\theta},z)}\right \rangle_{\theta z}/
  {r}
  \right).
 \end{aligned}
\end{equation} 
Here,
without regard to the temporal and spatial averaging processes, 
 the spatial distribution of $J_{\omega}$ is, 
\begin{equation}
 \begin{aligned}
  \mathit{J_{\omega}{(r,{\theta},z)}} 
  =r^2 {u_r}{(r,{\theta},z)} {u_{\theta}}{(r,{\theta},z)} - 
  \nu r^3 
  \partial_r
  \left(
  u_{\theta}{(r,{\theta},z)}/r
  \right).
 \end{aligned}
\end{equation}
When $J_{\omega}$ is divided by $r$, 
one could obtain the definition of angular velocity 
flux density 
\begin{equation}
 \begin{aligned}
  \mathit{q_{\omega}{(r,{\theta},z)}} & 
  =r{u_r}{(r,{\theta},z)} 
  {u_{\theta}}{(r,{\theta},z)} - 
  \nu r^2 
  \partial_r
  \left(
  u_{\theta}{(r,{\theta},z)}/{r}
  \right).
 \end{aligned}
\end{equation}
The convective part is 
$q^c_{\omega}=r{u_r}{(r,{\theta},z)} 
{u_{\theta}}{(r,{\theta},z)}$.
It is worth noting that, 
$\left \langle J_{\omega}\right \rangle_{\theta z}$ 
is the commonly conserved transverse current, 
whereas the density $\left \langle q_{\omega}\right \rangle_{\theta z}$
decreases along the radial direction as same as the heat flux density.

\subsection{\label{app:Effect} Effective viscosity $\nu_{t}/\nu$ and diffusivity 
$\kappa_{t}/\kappa$}

The global heat transfer could be defined as 
$Nu = { Q_{t} }/
{ Q_{t}^{Lam} }$, where 
$Q_{t}=\int_{V,\tau}q_{t}dVd\tau$ and
$Q_{t}^{Lam}=\int_{V,\tau}q_{t}^{Lam}dVd\tau$.
To describe the contribution of turbulent transport 
to the global heat transfer,  
the effective thermal diffusivity
$\kappa_{t}$ is defined as  
$Q_t=\int_{V,\tau}\kappa_{t}({\Delta}/{d})dVd\tau$.
Thus the dimensionless effective diffusivity is 
\begin{equation}
 \frac{\kappa_{t}}{\kappa}= aNu,
\end{equation}
where $a=2d^2/(ln(r_2/r_1)(r_2^2-r_1^2))$.

In the axially periodical domain or very long cylinders,  
$\left \langle J_{\omega}\right \rangle_{\theta z}$ 
remains constant radially.
However, owing to the braking 
effect of the fixed endwalls, 
$\left \langle J_{\omega}\right \rangle_{\theta z}$
decreases along the radial direction in our system. 
We consider the angular velocity flux at the inner cylinder
the $\omega$-Nusselt number for angular velocity transfer 
\cite{Leng_2021}

\begin{equation}
 Nu_{\omega}
 =\frac{\left \langle J_{\omega}\right \rangle_{\theta z, r=r_1}}
 {J_{\omega}^{Lam}}=
 \frac{u_1d}{2B} \frac{(R_1^2)(2Re_\tau)^2}{Re},
\end{equation}
where the friction Reynolds number $Re_\tau$ is defined as 
$Re_\tau=0.5u_\tau d/\nu$ 
with the friction velocity 
$u_\tau^2=-\nu r 
(\partial_r\left\langle 
u_{\theta}/r \right\rangle _{\theta z})$ at the inner wall.
Following Lathrop's estimation 
\cite{lathrop1992transition, 
 lathrop1992turbulent}, 
we define the effective viscosity (owing to the turbulent 
transport)
$\nu_{t}=G_{r_1}/(2\pi \rho u_{1}r_{1}^2h{d^{-1}})$, 
where the inner torque $G_{r_1}=2\pi r_{1}^3h\rho 
\nu  
{\partial_r}
\left( 
{\left\langle {u_{\theta}}/{r}\right\rangle_{\theta z}} 
\right)
= 2\pi r_{1}^2h {\rho 
 \nu^2}{d^{-2}} (2Re_\tau)^2$.
Thus, one could obtain the equation of the dimensionless effective 
viscosity \cite{Leng_2021}
\begin{equation}
 \frac{ \nu_{t}}{\nu}=\frac{(2Re_\tau)^2}{Re}.
\end{equation}

It is found that, the diffusivity  $\kappa_{t}/\kappa$ and 
dimensionless effective viscosity $\nu_{t}/\nu$  
have the same scaling with the global transport of heat ($Nu$) 
and angular momentum ($Nu_{\omega}$) respectively.

~\\

\nocite{*}
\providecommand{\noopsort}[1]{}\providecommand{\singleletter}[1]{#1}%
%


\end{document}